\documentclass{article}
\usepackage{spconf,amsmath,graphicx}

\usepackage{multirow}

\usepackage{caption}
\usepackage{subcaption}
\usepackage{comment}

\title{Front-End Adapter: Adapting Front-End Input of Speech based Self-Supervised Learning for Speech Recognition}

\name{
Xie Chen$^1$, Ziyang Ma$^1$, Changli Tang$^2$, Yujin Wang$^2$, 
Zhisheng Zheng$^1$}
\address{$^1$MoE Key Lab of Artificial Intelligence, AI Institute \\ X-LANCE Lab, Department of Computer Science and Engineering \\ Shanghai Jiao Tong University, Shanghai, China \\
$^2$Department of Electronic Engineering, Tsinghua University, Beijing, China}
\begin{document}
\ninept
\maketitle
\begin{abstract}
Recent years have witnessed a boom in self-supervised learning (SSL) in various areas including speech processing. Speech based SSL models present promising performance in a range of speech related tasks. However, the training of SSL models is computationally expensive and a common practice is to fine-tune a released SSL model on the specific task. It is essential to use consistent front-end input during pre-training and fine-tuning. This consistency may introduce potential issues when the optimal front-end is not the same as that used in pre-training.  In this paper,  we propose a simple but effective front-end adapter to address this front-end discrepancy. By minimizing the distance between the outputs of different front-ends,  the filterbank feature (Fbank) can be compatible with SSL models which are pre-trained with waveform. The experiment results demonstrate the effectiveness of our proposed front-end adapter on several popular SSL models  for the speech recognition task.
\end{abstract}
\section{Introduction}
\label{sec:intro}
Self-supervised learning (SSL) is attracting increasing research interests in deep learning for various fields including text, image and speech processing \cite{devlin2018bert, doersch2015imagessl}. The basic idea is to recover or predict itself based on its context information in an unsupervised fashion. Compared to text and image processing, it is not very straightforward to build a reconstruction or classification task for the speech signal due to its continuous character and strong short-time correlation. In recent years, significant research efforts have been made and promising progress achieved for speech-oriented SSL. Several representative SSL models are developed, such as Wav2vec \cite{baevski2020wav2vec2}, HuBERT \cite{hsu2021hubert}, WavLM \cite{chen2021wavlm} and Data2vec \cite{baevski2022data2vec}. Most of these SSL techniques are examined in the automatic speech recognition (ASR) task, where significant performance improvements are reported by using a small amount of supervised data given the SSL model pre-trained on large quantities of unlabeled audio. More recently, there are a series of works \cite{chen2021wavlm, yang2021superb, zhang2021bigssl, ma2022mt4ssl} demonstrating  that these SSL models are capable of extracting superior universal speech representations for various downstream speech-related tasks, e.g. speaker verification and emotion recognition.

Due to the continuous character of the speech signal, different types of
front-end can be used as input for SSL models, such as waveform and 
Mel-spectrum based features. In practice, the optimal 
front-end and stride size is task-dependent in light of accuracy and efficiency. 
However, the SSL models released by Fairseq \cite{ott2019fairseq} from 
Meta normally take waveform as the front-end input, and CNN blocks as the feature extractor, with a fixed stride size 
of 20ms, which is then fed into the Transformer based backbone model
during pre-training. In order to fully leverage the information learned 
from SSL training, it is essential to use the consistent front-end during 
pre-training, fine-tuning, and inference stages. 
However, the waveform based front-end is probably not the optimal
choice for all tasks, such as 
speech recognition. Therefore, this might limit potential 
applications of the publicly accessed speech based SSL models. It is 
of great practical value if we can lift this front-end consistency
constraint and allow the use of alternative front-ends to utilize the
power of SSL models in downstream tasks.

This paper aims to mitigate this front-end discrepancy issue
for speech based SSL models. More specifically, we explore the
feasibility of using Fbank feature in SSL models trained with 
waveform for the speech recognition task.
In addition to the standard ASR loss (e.g., CTC loss 
\cite{graves2006ctc}), another loss is introduced to minimize 
the distance of outputs from the waveform
and Fbank based front-ends during the fine-tuning stage. As a result,
the representation of the Fbank front-end can be adjusted towards that of the 
waveform front-end, so as to fit well with the rest of SSL model components. During the pre-training and fine-tuning stages, there is another potential mismatch with the stride size of input fed into the Transformer backbone model.
For example, in practice, the speech recognition system
might prefer to adopt a larger stride size (e.g. 40ms or 80ms)
instead of the default stride size used in the SSL model,
which is normally 20ms.
This stride size mismatch is also investigated in this paper. The experimental
results demonstrate that our proposed adapter is able to quickly adapt the SSL
model from the original waveform based front-end to the Fbank based front-end,  and comparable performances are achieved with different
front-end configurations.
To the best of our knowledge, this is the first work to explore the front-end
adapter for speech based SSL models.

This paper is organized as follows, Section \ref{sec:ssl} gives
a brief overview of three popular speech based SSL models, which are
Wav2vec 2.0 \cite{baevski2020wav2vec2}, HuBERT \cite{hsu2021hubert}
and Data2vec \cite{baevski2022data2vec} respectively.
Section \ref{sec:fea} presents the details of the proposed 
front-end adapter algorithm, followed by the experiments in 
Section \ref{sec:exps}. Finally, the conclusion is drawn in Section \ref{sec:conclusion}.

\section{Overview of Speech based Self-Supervised Learning}
\label{sec:ssl}
Due to the continuous character and strong short-time correlation lied
in the speech signal, the self-supervised learning of speech 
is not as straightforward as that of text and image. 
A range of algorithms are proposed to generate discrete or continuous representations from speech, so as to facilitate 
the  speech based self-supervised learning.
In this section, we will give a brief overview of three popular SSL models,
which are Wave2vec 2.0 \cite{baevski2020wav2vec2}, 
HuBERT \cite{hsu2021hubert} and Data2vec \cite{baevski2022data2vec}.
Figure \ref{fig:sslmodels} gives a diagram of these three SSL models.

\subsection{Wav2vec 2.0}
There are a series of continuous research efforts
\cite{schneider2019wav2vec, baevski2019vqwav2vec, baevski2020wav2vec2} 
aiming to convert the raw waveform into continuous and high-dimensional vectors, followed by the contrastive loss \cite{oord2018contrastiveloss} 
to distinguish the representation of the current input
from the competitive representations.
Wav2vec 2.0 is proven to be quite effective and provides an end-to-end
solution for self-supervised learning. The general pipeline of 
Wav2vec 2.0 could be illustrated as the green and blue blocks in Figure \ref{fig:sslmodels}.
In the Wav2vec 2.0 model \cite{baevski2020wav2vec2}, the output of the 
front-end processing block is quantified via product vector quantization or 
K-Means. Similar to the mask LM in BERT \cite{devlin2018bert}, 
the audio inputs are randomly  masked in a successive way and then
fed into Transformer blocks. The output of Wav2vec 2.0 aims to recover
the masked input by applying the contrastive loss, where the quantified
vector of the masked input is used as target and other quantified 
vectors from the codebook are randomly sampled as negative samples.

\subsection{HuBERT}
Instead of minimizing the reconstruction error with the learnable
codebook and contrastive loss in Wav2vec 2.0,
HuBERT \cite{hsu2021hubert} is proposed to assign a pseudo label for 
each speech frame. 
This pseudo label can be obtained from the K-means algorithm, 
on top of the standard MFCC features,
or the refined speech representations from a trained HuBERT model.
The resulting cluster ids from K-means
will be used as target for each frame 
and the corresponding input frames are masked to 
better utilize the context information and avoid information leakage.
HuBERT becomes increasingly popular and there are several
variants \cite{chen2021wavlm} for SSL developed on top of HuBERT.
Since each frame is associated with a pseudo label, cross
entropy could be simply used 
as the loss function and the training of 
HuBERT is found to be more stable compared to Wav2vec 2.0 with contrastive loss. 
The data flow of HuBERT can be depicted as the green
and yellow blocks in Figure \ref{fig:sslmodels}.

\subsection{Data2vec}
\begin{figure}[htbp]
\centering
\includegraphics[width=8cm]{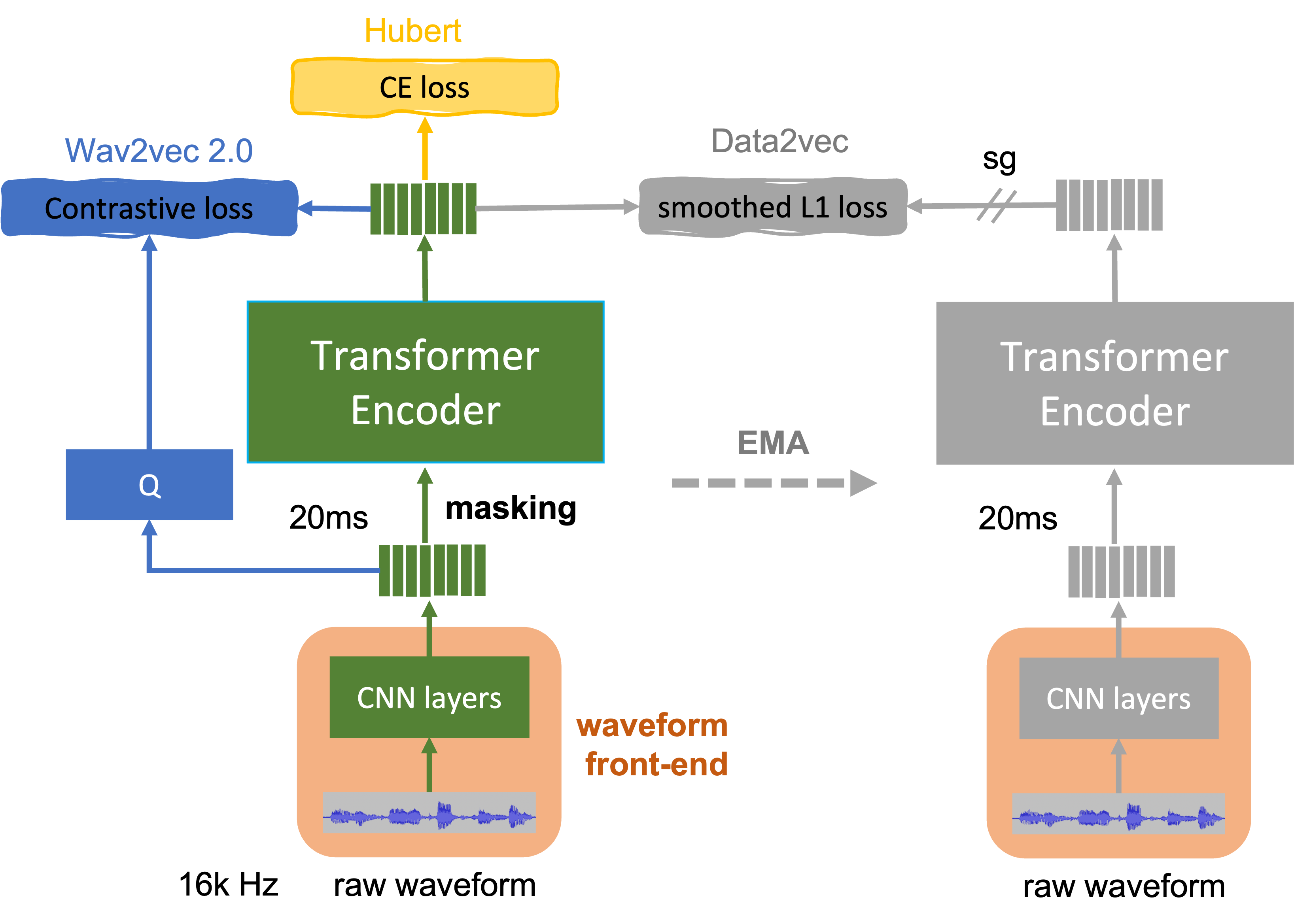}
\caption{Illustration of three different types of SSL models, 
the green block denotes the 
shared waveform front-end and Transformer backbone model. 
The blue part indicates the Wav2vec 2.0 model, the yellow part is
corresponding to HuBERT model, and the grey block is for Data2vec.}
\label{fig:sslmodels}
\end{figure}

In \cite{baevski2022data2vec}, Data2vec is proposed to unify
the SSL training algorithm for different modalities, including audio,
image and text. Two parallel models are employed to serve as teacher and 
student models respectively, at every training step, the teacher model 
is allowed to see the complete input, while the student model consumes 
the masked input. The objective function is to minimize the 
distance between the outputs of teacher and student models 
in the masked regions, where the smoothed L1 loss is applied. 
The gradient computed from the loss function is only back-propagated 
to the student model and the teacher model is a delayed version of
the student model by applying the EMA  technique \cite{grill2020byol}.
At the end of training, the teacher model is discarded and the 
student model is retained as the resulting SSL model. The green and grey 
blocks in Figure \ref{fig:sslmodels} shows the general 
framework of the Data2vec model.

After the pre-training of SSL models on the unlabeled audio data,
the front-end processing block and 
Transformer layers, which are corresponding to the green blocks in Figure
\ref{fig:sslmodels}, are kept as the acoustic encoder, 
and then fine-tuned jointly with the rest of task-dependent model components
on downstream tasks, e.g. speech recognition, speaker verification. 
There are multiple choices for the front-end input in speech processing,
e.g. waveform and Mel-spectrum feature. The above-mentioned three 
SSL models take waveform as the front-end input in the original papers.
The code for SSL model pre-training and fine-tuning is released
in the Fairseq toolkit \footnote{https://github.com/facebookresearch/fairseq}
from Meta. In addition, a range of well-trained
SSL models from Meta are also provided and can be publicly accessed. These models 
are widely used in the community for various downstream tasks.  In all models 
released by Fairseq, waveform is chosen as the  front-end input, and 
waveform based front-end consisting of several CNN layers yields the acoustic 
feature representation with a stride size of 20ms, as shown in Figure \ref{fig:sslmodels}.
The front-end input of speech should be consistent 
during pre-training, fine-tuning and inference stages.
This means that if we want to fully leverage the power 
of the existed SSL models, we need to use the same speech front-end.
However, in many existed solutions or production systems, 
waveform might not be the optimal choice in light of accuracy and 
efficiency. For instance, in the ASR task, it is preferred to adopt
Fbank feature as input and a larger stride size, e.g. 40ms, to
reduce computation in test time.
Hence, the demand of the front-end input consistency between 
pre-training and fine-tuning, might limit potential applications
of these released SSL models. In this paper, we aim to mitigate this mismatch 
and investigate the use of Fbank feature during fine-tuning on the SSL
models originally pre-trained with the waveform as front-end input, which is detailed in the next section.

\section{Front-End Adapter for SSL models}
\label{sec:fea}

As mentioned in the previous section, it is computationally expensive
for the pre-training of SSL models, a common practice is downloading a publicly released
SSL model and then fine-tune it with a small amount of labeled data
on downstream tasks.  In this paper, we mainly investigate the speech
recognition task as the downstream task for the experiment.
It is necessary to use the consistent front-end input 
during the pre-training and fine-tuning stages,
where the waveform is the default front-end input for various SSL
models from Fairseq. 
Due to the variants of the front-end input existed
in the speech signal, people might prefer to use different
front-end input rather than waveform, e.g. Fbank feature, in the 
practical scenario. Hence, this
discrepancy brings about potential challenges to fully utilize the 
existing and well-trained SSL models. In this section, 
we propose a simple and effective framework to adapt the front-end, which allows different front-ends to be used during pre-training and fine-tuning.

\begin{figure}[htbp]
     \centering
     \begin{subfigure}[b]{0.4\textwidth}
         \centering
         \includegraphics[width=0.8\textwidth]{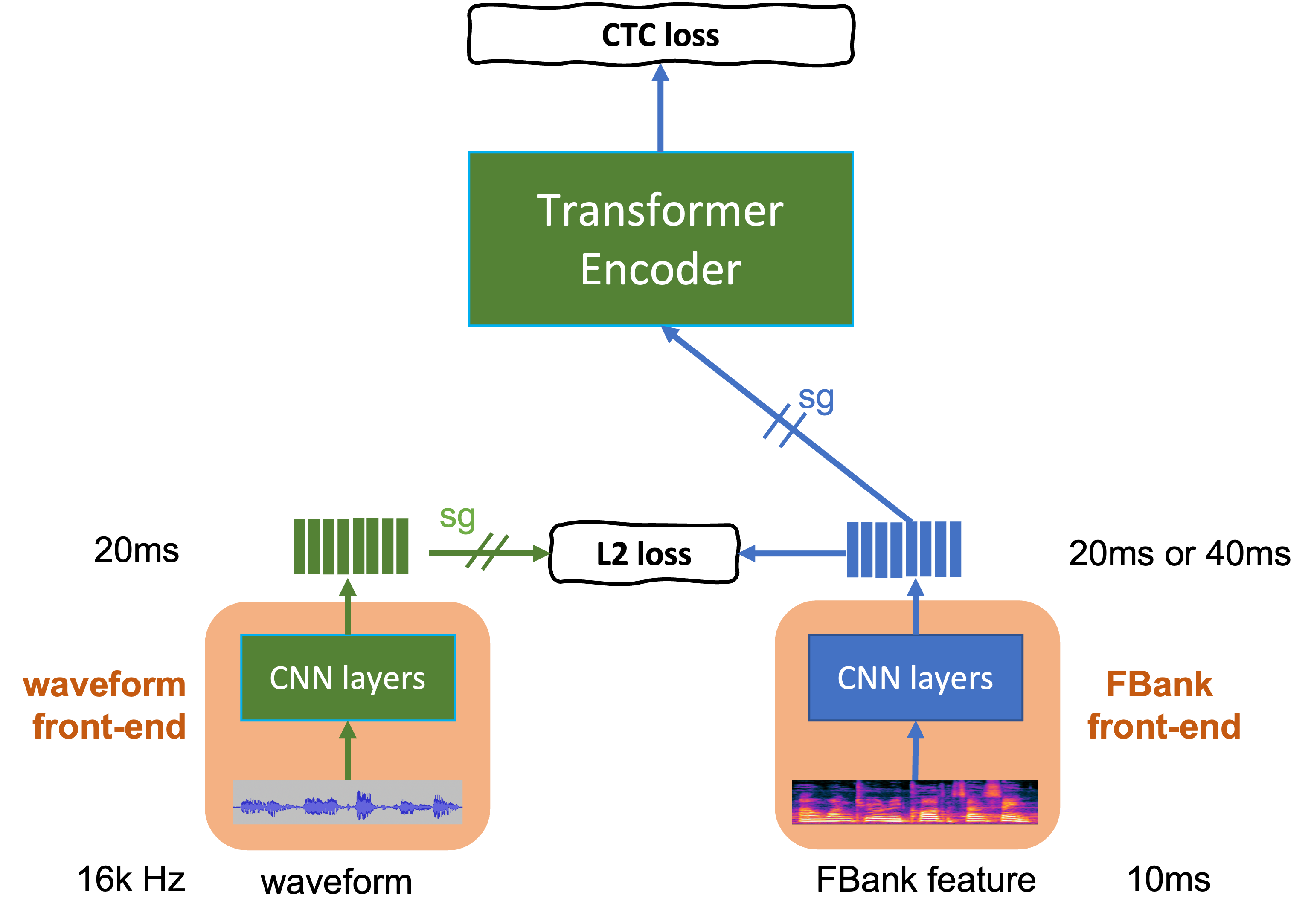}
         \caption{\textbf{Stage 1} of the front-end adapter:
         both CTC loss and L2 loss are applied, and the L2 loss won't back-propagate
         to the waveform front-end block and CTC loss won't back-propagate to the
         Fbank front-end during stage 1.}
         \label{fig:fea_1}
     \end{subfigure}
     \hfill
     \begin{subfigure}[b]{0.4\textwidth}
         \centering
         \includegraphics[width=0.4\textwidth]{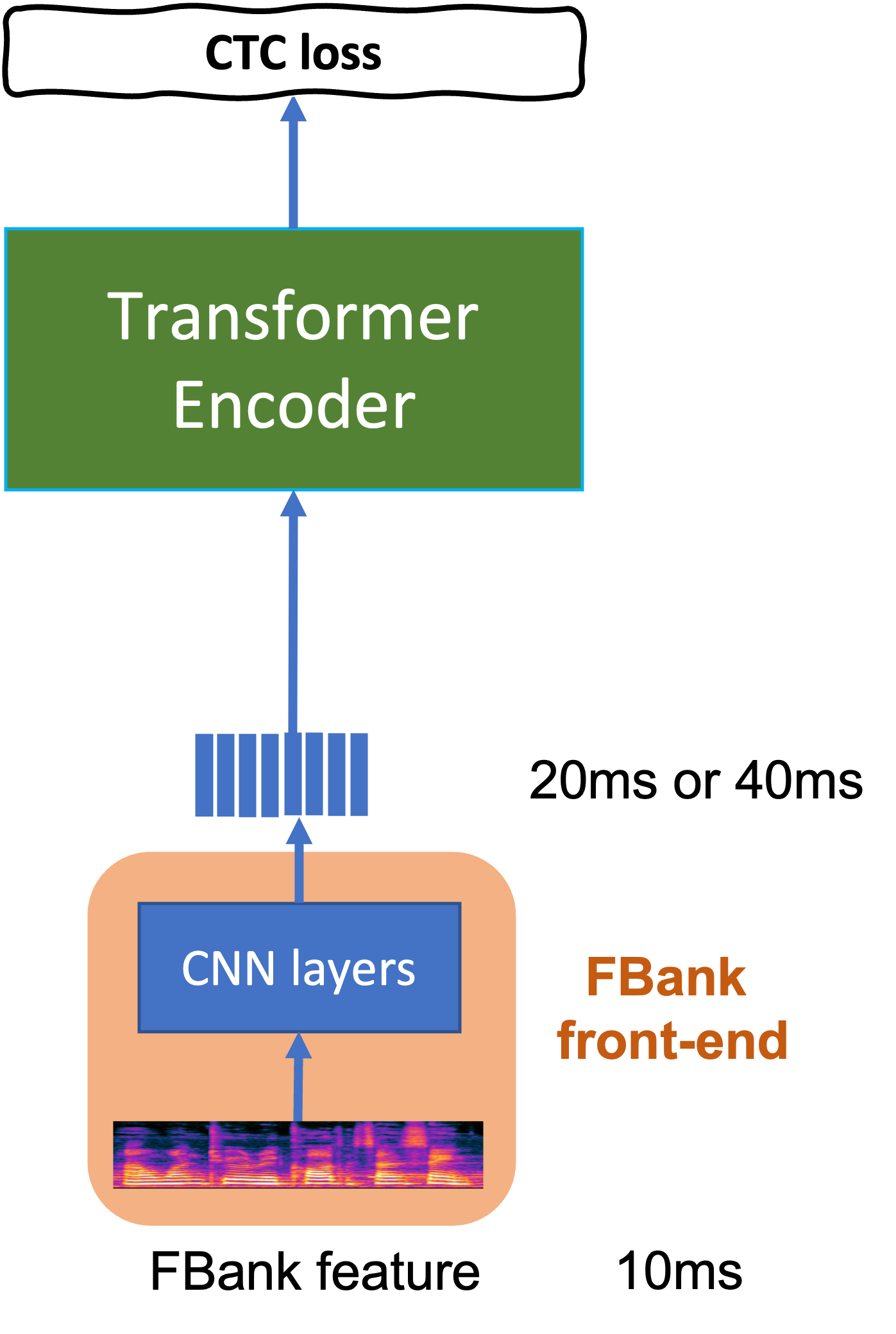}
         \caption{\textbf{Stage 2} of the front-end adapter: all modules are jointly
         optimized with CTC loss}
         \label{fig:fea_2}
     \end{subfigure}
        \caption{Two stages of the front-end adapter for SSL models}
        \label{fig:fea}
\end{figure}

As shown in Figure \ref{fig:sslmodels}, the green blocks (e.g. CNN layers
front-end and Transformer layers) will be kept after pre-training and then
fine-tuned with labeled data. More specifically, in speech recognition, a
CTC output layer can be added to the output of Transformer layers.
According to our early attempts, simply replacing the waveform front-end
with a randomly initialized Fbank front-end did not converge well. 
It indicates that the mismatch lied in different front-ends is detrimental for
the well-trained SSL model.
In order to address this potential issue, a novel front-end adapter 
is proposed to alleviate the gap between the waveform and Fbank based 
front-ends.
Figure \ref{fig:fea} illustrates the general process of the proposed
front-end adapter. The Fbank feature based front-end 
shown in the right bottom is used to replace the original waveform front-end.
However, the straightforward replacement of front-end input does not
work well due to the differences lied in different front-end processing 
blocks. To resolve this mismatch, the output of the original waveform front-end
is used to regularize the Fbank front-end output, which is shown in the
bottom blocks of Figure \ref{fig:fea}.
The original waveform front-end is frozen during the training and is
used to regularize the Fbank front-end output. An L2 loss is applied
to minimize the distance between outputs of the waveform and Fbank
based front-ends. 
It is noteworthy that this additional L2 loss is only applied in a 
specified number of steps at the early stage of training, which is denoted as the adapter warm-up stage.
In this paper, we set the adapter warm-up stage to the first 3 epochs
as an empirical value. 
Furthermore, we find during the adapter warm-up stage, 
when the CTC loss and L2 loss are both used to update the Fbank
front-end block, the gradient back-propagated from the CTC loss may
overwhelm the gradient of L2 loss. Hence. In order to avoid this effect,
during the adapter warm-up stage, 
the gradient from the CTC loss is only used to update the Transformer 
and CTC layers, not used to update the Fbank front-end block. The L2 loss between the waveform and Fbank front-ends is applied to update the Fbank front-end.
After the adapter warm-up stage, the Fbank front-end, Transformer and 
CTC layers are jointly optimized with the CTC loss as standard CTC based
ASR training.
In summary, the loss function can be written as below. 
    \[
    \mathcal{L} = 
\begin{cases}
    \mathcal{L}_{ctc} + \mathcal{L}_{l2},& \text{if } n\le N_w\\
     \mathcal{L}_{ctc},              & \text{otherwise}
\end{cases}
\]

Another important factor to consider is the ability to use variable
stride sizes for the output of different front-end blocks. For the SSL
models released in Fairseq, the stride size of the waveform is normally
down-sampled to 20ms. In contrast, a 
larger stride size, e.g. 40ms, is preferred for efficiency
consideration in practice. 
This will cause the length mismatch between the outputs of the 
waveform and Fbank front-ends. In order to resolve this 
issue, we simply down-sample the output of the waveform front-end
to match with the Fbank front-end, e.g. 40ms, and then compute the L2 distance.

Overall, the fine-tuning of the SSL model with the proposed front-end adapter can be 
carried out in two stages.
\begin{enumerate}
    \item Front-end adapter warm-up: when the update step is smaller than
    the warm-up step (e.g. the first 3 epochs in this paper),
    both CTC loss and L2 loss are applied. Note that the gradient of the L2 loss 
    is not used for back-propagation to the waveform front-end block
    and the gradient of CTC loss won't be back-propagated to the Fbank front-end
    in this stage.
    \item Fine-tuning: after the adapter warm-up stage, all modules then is 
    fine-tuned with the standard CTC loss.
\end{enumerate}

It is worth mentioning that the front-end adapter framework can 
be easily extended to those three SSL models, i.e. Wav2vec 2.0, HuBERT and Data2vec, introduced in Section \ref{sec:ssl}
since only the waveform front-end and Transformer layers 
in Figure \ref{fig:sslmodels} are needed during the front-end adapter and the following
fine-tuning stage.

\section{Experiments}
\label{sec:exps}


In this section, three popular SSL models,
Wav2vec 2.0, HuBERT and Data2vec, from the Fairseq
repository \footnote{https://github.com/facebookresearch/fairseq} 
are investigated. All of these three SSL models are 
pre-trained on the 960-hour Librispeech data with ``base" model 
configuration, and the number of parameters for these models are
comparable with around 97M. 
A subset of two public English data sets, Librispeech \cite{panayotov2015librispeech} and Gigaspeech \cite{chen2021gigaspeech},
are used in the downstream speech recognition task to validate
the effectiveness of the proposed front-end adapter in Section
\ref{sec:fea}. For Gigaspeech, a 100-hour speech is randomly 
selected as fine-tuning data and the standard dev and test 
data are used for performance evaluation; while for Librispeech,
the 100-hour train-clean speech is chosen as fine-tuning data, and 
the dev-other and test-other are used for evaluation. Letter is 
used as the model unit as the output sequence for speech recognition.
The CTC criterion is used as the loss function for ASR training. 
SpecAug \cite{park2019specaugment} is 
applied in the Fbank feature front-end to improve the performance of speech
recognition, and the default fine-tuning configuration from Fairseq is
adopted for the waveform front-end \footnote{The random masks 
applied in the output of the waveform front-end during 
fine-tuning have a similar effect as SpecAug}.
The WER results from greedy search without any LM are reported for
performance evaluation for simplicity.

\begin{figure}[htbp]
\centering
\includegraphics[width=7cm]{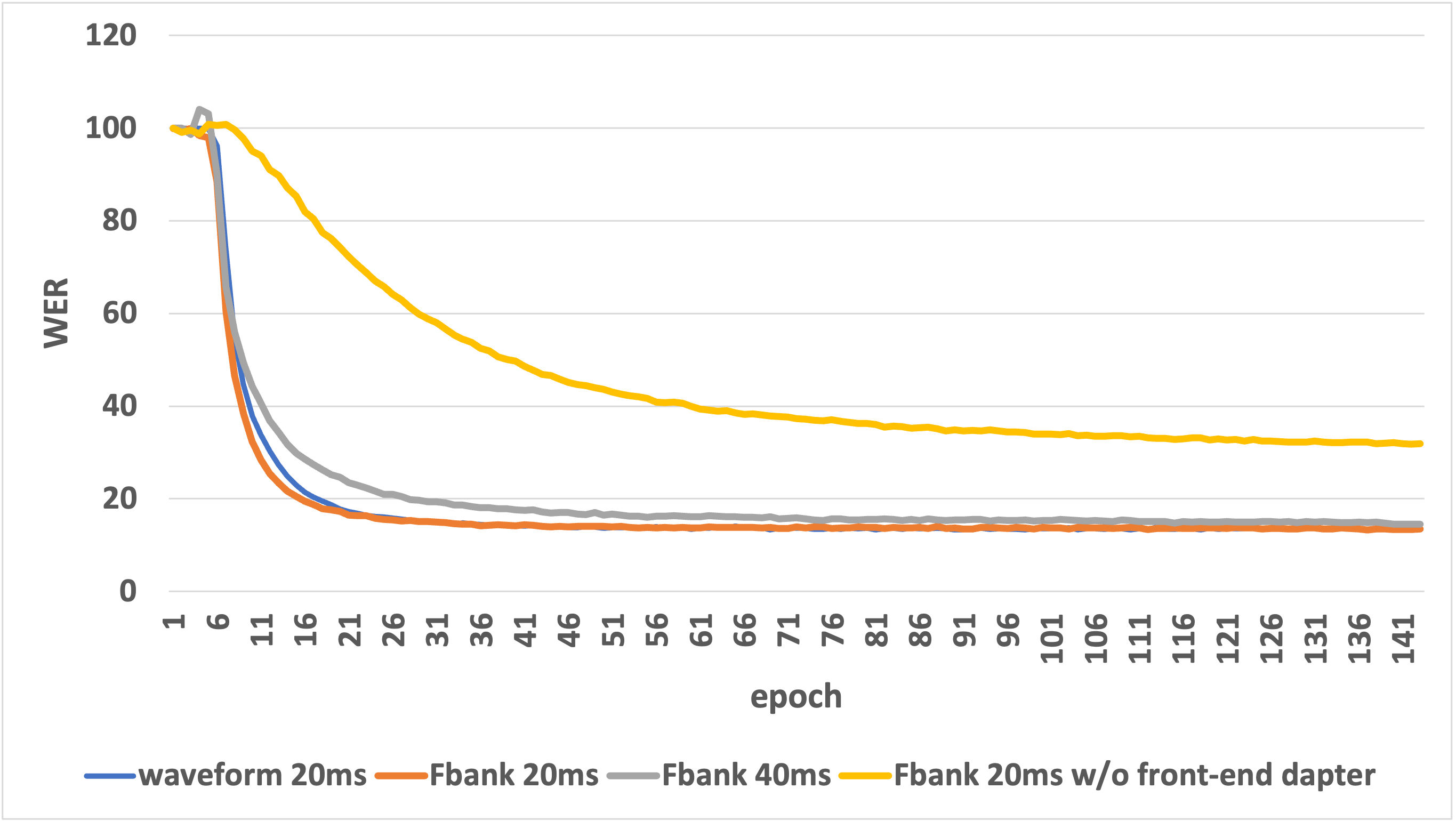}

\caption{Fine-tuning WER results of Waveform, Fbank based front-ends with different stride sizes for HuBERT. The SSL models are fine-tuned on 100-hour Librispeech data and WERs are reported on the dev-other set and }
\label{fig:wer}
\end{figure}

The first experiment is to investigate the effect of the proposed 
front-end adapter on the 100-hour subset of Librispeech. 
The HuBERT model is chosen as the SSL model for experimental analysis.
Figure \ref{fig:wer} plots the WER trends during fine-tuning with
different front-end and stride size. 
It can be seen that the waveform based front-end (blue curve)
converges quickly, which is expected as the same front-end is used during pre-training and fine-tuning. 
The adapted Fbank front-end with the sample stride size (i.e. 20ms, colored as orange) achieves similar performances compared to HuBERT on the waveform. 
The adapted Fbank front-end with a stride size of 40ms (grey curve)
also converges quickly and yields a slightly worse performance
compared to that of 20ms stride size. This degradation could be explained by the limited amount of fine-tuning data. The yellow curve plots the WER trend during fine-tuning when simply replacing the waveform front-end by a Fbank front-end without adapting front-end. It can be seen that without the front-end adapter, the mismatch existed in the Fbank front-end and Transformer encoder degrades the fine-tuning WER performance significantly. Hence, it is critical to employ the front-end adapter during fine-tuning when different front-end is used.

\begin{figure}[htbp]
\centering
\includegraphics[width=6cm]{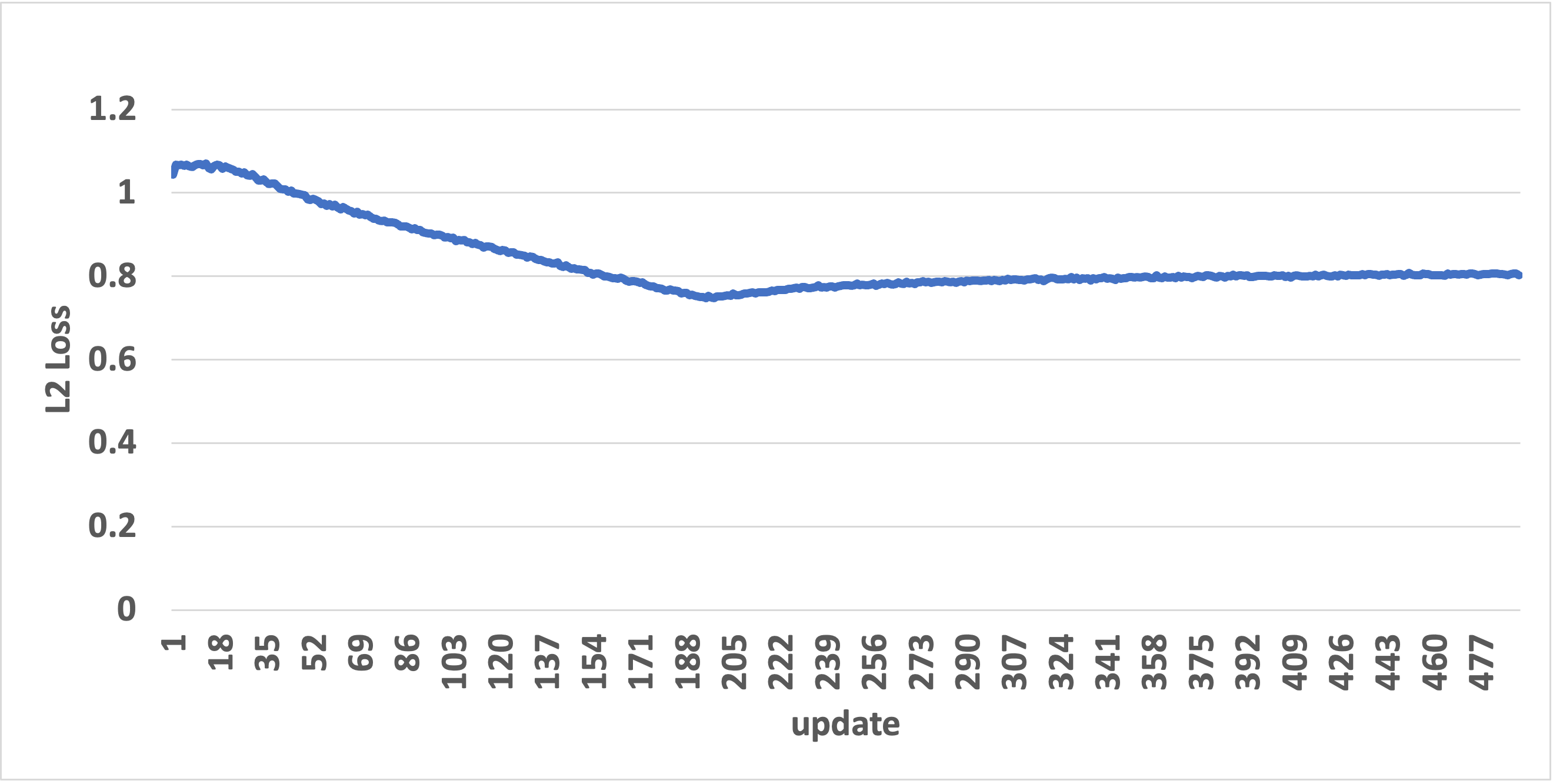}

\caption{The Euclidean distance between waveform and Fbank front-ends during training. The first 200 updates are used for Stage 1 of the front-end adapter.}
\label{fig:l2loss}
\end{figure}

Figure \ref{fig:l2loss} also plots the change of Euclidean distance between the outputs
of waveform and Fbank based front-ends. It can be seen that during 
 stage 1 of the front-end adapter (the first 200 updates), the Euclidean distance keeps reducing due
to the additional L2 loss.
In stage 2 of the front-end adapter (after the first 200 updates), there is no L2 loss
applied in the objective function, the Euclidean distance remains stable. After Stage 1, the Fbank front-end can be adjusted to fit well with the Transformer encoder.

Table \ref{tab:WERonthreesslmodels} reports the WER results 
after fine-tuning on 
the three types of SSL models with different front-ends
(Fbank vs waveform) and stride sizes (20ms vs 40ms).
It could be 
seen that our front-end adapter could effectively adapt the 
waveform based front-end to Fbank based front-end. For Wav2vec 2.0 and HuBERT model, the Fbank front-end with a stride size of 20ms outperform the waveform front-end, and the Fbank with a stride size of 40ms yield comparable performance to the waveform front-end. In Data2vec, the waveform based front-end gives the best WER among the three configurations. Nevertheless, the Fbank front-end model adapted from Data2vec still performs better than all models fine-tuned on Wav2vec 2.0 and HuBERT. 

\begin{table}[htbp]
    \caption{WER results on the Librispeech and Gigaspeech test sets on three SSL models. The standard dev and test sets are used for Gigaspeech, and the dev-other and test-other sets are adopted for evaluation. }
    \centering
    \begin{tabular}{|c|c|c||c|c|c|c|c|c|}
    \hline
   \multirow{1}{*}{SSL }  & Front-end & Stride & \multicolumn{4}{|c|}{WER} \\
    \cline{4-7}
       model &    input     & size   &   \multicolumn{2}{|c|}{Gigaspeech} & \multicolumn{2}{|c|}{Librispeech} \\
                & & (ms)& dev & test & dev & test \\
                
    \hline\hline
        \multirow{3}{*}{Wav2vec2}
                    & Wave & 20 &  20.8 &20.3           & 14.29&14.54      \\    
                    & Fbank  & 20 &  \bf{20.0} & \bf{19.8}     & \bf{14.21} & \bf{14.13}      \\
                    & Fbank  & 40 &  20.3 & 20.3     & 15.28   &  15.15 \\  
    \hline  
        \multirow{3}{*}{HuBERT}      
                    & Wave & 20 & 20.2 & 19.9  & 14.39 & 14.12      \\   
                    & Fbank  & 20 &  \bf{19.2} & \bf{19.3}  &  \bf{13.73} & \bf{13.79}     \\ 
                    & Fbank  & 40 &  19.8 & 19.9    & 14.56   & 14.47 \\
    \hline
        \multirow{3}{*}{Data2vec}
                    & Wave & 20 & \bf{18.3} & \bf{18.1}     & \bf{9.99} & \bf{10.13}       \\    
                    & Fbank  & 20 &  18.8 & 19.1    &  11.11 & 11.25     \\ 
                    & Fbank & 40  &  20.1 & 20.3   & 12.67  & 12.78  \\ 
    \hline
    \end{tabular}

    \label{tab:WERonthreesslmodels}
    \vspace{-0.1cm}
\end{table}

\section{Conclusion and Discussion}
\label{sec:conclusion}
In recent years, speech based self-supervised 
learning (SSL) presents
promising performance improvements on a range
of downstream tasks 
including speech recognition. 
Nowadays, most popular and publicly released SSL
models adopt waveform as the front-end input. 
In order to leverage the power of SSL models, it is 
necessary to use consistent front-end input during 
pre-training and fine-tuning stages, which might limit
the potential applications of SSL 
models as the waveform is not the optimal choice in many 
tasks in practical scenarios. 
In this paper, we proposed a simple and effective front-end
adapter framework to mitigate the mismatch between
different front-end inputs for speech based 
SSL models. Our proposed front-end adapter allows Fbank 
feature to be used in the SSL models originally trained
on the waveform, and retains comparable performance in the
speech recognition task. 


\section{Acknowledgement}
This work was supported by the National Natural Science Foundation of China under Grant No. 6220070337, and the International Cooperation Project of PCL.

\vfill\pagebreak
\bibliographystyle{IEEEbib}
\bibliography{refs}

\begin{thebibliography}{10}

\bibitem{devlin2018bert}
Jacob Devlin, Ming-Wei Chang, Kenton Lee, and Kristina Toutanova,
\newblock ``{BERT}: Pre-training of deep bidirectional {T}ransformers for
  language understanding,''
\newblock {\em arXiv preprint arXiv:1810.04805}, 2018.

\bibitem{doersch2015imagessl}
Carl Doersch, Abhinav Gupta, and Alexei Efros,
\newblock ``Unsupervised visual representation learning by context
  prediction,''
\newblock in {\em IEEE Proc. ICCV}, 2015, pp. 1422--1430.

\bibitem{baevski2020wav2vec2}
Alexei Baevski, Yuhao Zhou, Abdelrahman Mohamed, and Michael Auli,
\newblock ``wav2vec 2.0: A framework for self-supervised learning of speech
  representations,''
\newblock {\em Proc. NIPS}, vol. 33, pp. 12449--12460, 2020.

\bibitem{hsu2021hubert}
Wei-Ning Hsu, Benjamin Bolte, Yao-Hung~Hubert Tsai, Kushal Lakhotia, Ruslan
  Salakhutdinov, and Abdelrahman Mohamed,
\newblock ``{HuBERT}: Self-supervised speech representation learning by masked
  prediction of hidden units,''
\newblock {\em IEEE/ACM TALSP}, vol. 29, pp. 3451--3460, 2021.

\bibitem{chen2021wavlm}
Sanyuan Chen, Chengyi Wang, Zhengyang Chen, Yu~Wu, Shujie Liu, Zhuo Chen, Jinyu
  Li, Naoyuki Kanda, Takuya Yoshioka, Xiong Xiao, et~al.,
\newblock ``{WavLM}: Large-scale self-supervised pre-training for full stack
  speech processing,''
\newblock {\em arXiv preprint arXiv:2110.13900}, 2021.

\bibitem{baevski2022data2vec}
Alexei Baevski, Wei-Ning Hsu, Qiantong Xu, Arun Babu, Jiatao Gu, and Michael
  Auli,
\newblock ``Data2vec: A general framework for self-supervised learning in
  speech, vision and language,''
\newblock {\em arXiv preprint arXiv:2202.03555}, 2022.

\bibitem{yang2021superb}
Shu-wen Yang, Po-Han Chi, Yung-Sung Chuang, Cheng-I~Jeff Lai, Kushal Lakhotia,
  Yist~Y Lin, Andy~T Liu, Jiatong Shi, Xuankai Chang, Guan-Ting Lin, et~al.,
\newblock ``{SUPERB}: Speech processing universal performance benchmark,''
\newblock {\em arXiv preprint arXiv:2105.01051}, 2021.

\bibitem{zhang2021bigssl}
Yu~Zhang, Daniel~S Park, Wei Han, James Qin, Anmol Gulati, Joel Shor, Aren
  Jansen, Yuanzhong Xu, Yanping Huang, Shibo Wang, et~al.,
\newblock ``{BigSSL}: Exploring the frontier of large-scale semi-supervised
  learning for automatic speech recognition,''
\newblock {\em arXiv preprint arXiv:2109.13226}, 2021.

\bibitem{ma2022mt4ssl}
Ziyang Ma, Zhisheng Zhen, Changli Tang, Yujin Wang, and Xie Chen,
\newblock ``Mt4ssl: Boosting self-supervised speech representation learning by
  integrating multiple targets,''
\newblock {\em arXiv preprint arXiv:2211.07321}, 2022.

\bibitem{ott2019fairseq}
Myle Ott, Sergey Edunov, Alexei Baevski, Angela Fan, Sam Gross, Nathan Ng,
  David Grangier, and Michael Auli,
\newblock ``{FAIRSEQ}: A fast, extensible toolkit for sequence modeling,''
\newblock {\em arXiv preprint arXiv:1904.01038}, 2019.

\bibitem{graves2006ctc}
Alex Graves, Santiago Fern{a}ndez, Faustino Gomez, and J{u}rgen Schmidhuber,
\newblock ``Connectionist temporal classification: labelling unsegmented
  sequence data with recurrent neural networks,''
\newblock in {\em Proc. ICML}, 2006, pp. 369--376.

\bibitem{schneider2019wav2vec}
Steffen Schneider, Alexei Baevski, Ronan Collobert, and Michael Auli,
\newblock ``wav2vec: Unsupervised pre-training for speech recognition,''
\newblock {\em arXiv preprint arXiv:1904.05862}, 2019.

\bibitem{baevski2019vqwav2vec}
Alexei Baevski, Steffen Schneider, and Michael Auli,
\newblock ``{VQ}-wav2vec: Self-supervised learning of discrete speech
  representations,''
\newblock {\em arXiv preprint arXiv:1910.05453}, 2019.

\bibitem{oord2018contrastiveloss}
Aaron van~den Oord, Yazhe Li, and Oriol Vinyals,
\newblock ``Representation learning with contrastive predictive coding,''
\newblock {\em arXiv preprint arXiv:1807.03748}, 2018.

\bibitem{grill2020byol}
Jean-Bastien Grill, Florian Strub, Florent Altch{\'e}, Corentin Tallec, Pierre
  Richemond, Elena Buchatskaya, Carl Doersch, Bernardo Avila~Pires, Zhaohan
  Guo, Mohammad Gheshlaghi~Azar, et~al.,
\newblock ``Bootstrap your own latent-a new approach to self-supervised
  learning,''
\newblock {\em Advances in neural information processing systems}, vol. 33, pp.
  21271--21284, 2020.

\bibitem{panayotov2015librispeech}
Vassil Panayotov, Guoguo Chen, Daniel Povey, and Sanjeev Khudanpur,
\newblock ``Librispeech: an asr corpus based on public domain audio books,''
\newblock in {\em IEEE Proc. ICASSP)}, 2015, pp. 5206--5210.

\bibitem{chen2021gigaspeech}
Guoguo Chen, Shuzhou Chai, Guanbo Wang, Jiayu Du, Wei-Qiang Zhang, Chao Weng,
  Dan Su, Daniel Povey, Jan Trmal, Junbo Zhang, et~al.,
\newblock ``Gigaspeech: An evolving, multi-domain asr corpus with 10,000 hours
  of transcribed audio,''
\newblock {\em arXiv preprint arXiv:2106.06909}, 2021.

\bibitem{park2019specaugment}
Daniel~S Park, William Chan, Yu~Zhang, Chung-Cheng Chiu, Barret Zoph, Ekin~D
  Cubuk, and Quoc~V Le,
\newblock ``Specaugment: A simple data augmentation method for automatic speech
  recognition,''
\newblock {\em arXiv preprint arXiv:1904.08779}, 2019.

\end{thebibliography}

\end{document}